\documentclass[twocolumn,showpacs,preprintnumbers,amsmath,amssymb,prl]{revtex4}
\usepackage{amsmath}
\usepackage{amsfonts}
\usepackage{amssymb}
\usepackage{graphicx}
\usepackage{psfig,epsfig}
\def\tn{\textnormal}

\begin{document}

\title{Shear instabilities in granular mixtures} 
\author{Massimo Pica Ciamarra}
\email{picaciam@na.infn.it}
\author{Antonio Coniglio}
\email{coniglio@na.infn.it}
\author{Mario Nicodemi}
\email{mario.nicodemi@na.infn.it}
\homepage{http://smcs.na.infn.it}
\affiliation{Dip.to di Scienze Fisiche, Universit\`{a} di Napoli ``Federico II'' \\
INFM-Coherentia, INFN and AMRA, Napoli, Italy}
\pacs{45.70.Qj, 45.70.Mg, 47.54.+r}

\date{December 21, 2004}

\begin{abstract}
Dynamical instabilities in fluid mechanics are responsible of a variety of important common phenomena,
such as waves on the sea surface or Taylor vorteces in Couette flow.
In granular media dynamical instabilities has just begun to be discovered.
%\cite{Shinbrot2001,Goldfarb2002,Shinbrot2004,Jaeger1996}.
Here we show by means of molecular dynamics simulation the existence of a new dynamical 
instability of a granular mixture under oscillating horizontal shear, which leads to the formation of a striped pattern where the components are segregated. 
We investigate the properties of such a Kelvin-Helmholtz like instability and 
show how it is connected to pattern formation in granular flow and 
segregation.
%\cite{Mullin2000,Reis2002,Yeomans,King}.
\end{abstract}

\maketitle
A simple example of instabilities in fluid mechanics is the Kelvin-Helmholtz instability where a flat interface between two fluids flowing one past the other at different velocities is unstable, explaining for instance why waves on the see surface form~\cite{Chandrasekhar1961,Swinney}. %Lamb1932,
In athermal systems such as granular mixtures, where an hydrodynamic like theory is not yet well established, the discovery of instabilities is a recent achievement \cite{Shinbrot2001,Goldfarb2002,Shinbrot2004,Jaeger1996}. 
%By studying a binary mixture of grains on an horizontal oscillating plate we report here of a new dynamical instability of a granular mixture.
By studying the evolution of two bands of different grains placed on an horizontally oscillating tray (see Fig. 1(a)), we discover a new dynamical instability. This instability appears as growing wavy interface, which, at long times, leads
to a pattern of alternating segregated stripes of grains, perpendicular to the driving direction.
We suggest that this type of surface shear instability is the common mechanism for apparently different phenomena such 
wave formation as found in experiments on granular flow \cite{Goldfarb2002}, and segregation processes \cite{segre} as observed
 in \cite{Mullin2000,Reis2002,Reis2004,King}, initially explained in terms of a thermodynamic driven phase separation.
%as stripe formation starting from a mixed configuration (as already found in \cite{Mullin2000,Reis2002,Reis2004,King}) and wave formation as found in experiments on granular flow \cite{Goldfarb2002}, and reproduced by our simulations. 
In our simulations we determine the region of existence of the pattern formation - segregation process as a function of the area fractions of large and small grains, and we discuss the dependence of the pattern as a function of the amplitude and frequency of oscillation.

%We study the behaviour of a granular mixture of disks on an horizontal oscillating
%plate by performing Molecular Dynamics (MD) simulations (also known as Discrete Element Methods), 
%of a binary system of disks which lay on a tray (i.e., in two dimensions) with periodic 
%boundary conditions in the x-direction and hard walls in the other. Grains interact with the 
%tray by a viscous force proportional to their relative velocity via a viscosity parameter $\mu$ different for the two species; when overlapping, grains interact via 2D Hertzian contact forces
%\cite{herrmann_luding}. Simulations details and parameters, chosen to model experimental conditions
%similar to those investigated by Mullin and coworkers
%\cite{Mullin2000,Reis2002,Reis2004,Ehrhardt2004}, 
%are given at the end of the paper.

%\textit{Simulation details --}
\textit{MD simulation model --}
We perform Molecular Dynamics (MD) simulations (also known as Discrete Element Methods)
of a binary system of disks which lay on a tray (i.e., in two dimensions) with periodic 
boundary conditions in the x-direction and hard walls in the other. Grains interact with the 
tray by a viscous force proportional to their relative velocity via a viscosity parameter $\mu$ different for the two species; when overlapping, grains interact via 2D Hertzian contact forces
\cite{herrmann_luding}. Simulations details and parameters are chosen to model experimental conditions similar to those investigated by Mullin and coworkers \cite{Mullin2000,Reis2002,Reis2004,Ehrhardt2004}.

Two grains with diameters $D_i$ and $D_j$ in positions ${\bf r}_i$ and ${\bf r}_j$ interact if overlapping, i.e., if
$\delta_{ij} = [(D_i + D_j)/2 - |{\bf r}_i-{\bf r}_j|] > 0$. 
The interaction is given by a normal Hertz force with viscous dissipation \cite{Cundall,Silbert2001}.
In 2D this reduces to the linear spring-dashpot model, 
${\bf f}_n = k_n\delta_{ij}{\bf n}_{ij} - \gamma_n m_\textnormal{red}{\bf v}_{nij}$,
where $k_n$ and $\gamma_n$ are the elastic and viscoelastic constants,
and $m_\textnormal{red}=m_im_j/(m_i+m_j)$ is the reduced mass.
As in \cite{Ehrhardt2004} we model the interaction with the tray via a viscous force
$\bf{f}_t = -\mu ({\bf v}-{\bf v}_\tn{tray})$,
where ${\bf v}_\tn{tray}(t) = 2\pi A \nu \sin(\nu t) {\bf x}$
is the velocity of the tray and ${\bf v}$ the velocity of the disk,
plus a white noise force ${\bf \xi}(t)$ with
$\langle {\bf \xi}(t) {\bf \xi}(t') \rangle = 2\Gamma \delta(t-t')$, where $\Gamma = 0.2$ g$^2$cm$^2$s$^{-3}$.
The interaction between particles and walls is elastic.
We solve the equations of motion by Verlet algorithm with an time-step of $6\mu$s.
For the grain-grain interaction, we use the value
$k_n = 2\;10^5$ g cm$^2$s$^{-2}$ and $\gamma_n$ chosen,
for each kind of grains, such that the restitution coefficient
is given: $e=0.8$ \cite{Silbert2001}.
The two components of our mixture (named $b$ and $s$) have
%mass $M_b = 1$ g and $M_s = 0.03$ g, and
viscous coefficient $\mu_b = 0.28$ g s$^{-1}$ and $\mu_s = 0.34$ g s$^{-1}$.
%In studying a mixture of equal sized grains we use $D_b = D_s = 1$ cm.
%Otherwise case $D_b = 1$ cm and
Apart from a simple rescaling of masses and lengths, these values are those
of Ref. \cite{Ehrhardt2004} (and given in private communications),
and are taken from direct measurements on the experimental system
in \cite{Mullin2000}.
The size of the tray is $L_x = 320$ cm, $L_y = 16$ cm.
The qualitative picture we discuss does not change if these values are
changed.

\textit{Results --}
We consider by now a mixture of heavy grains, of mass $M_b=1$g and area fraction $\phi_b=0.37$, %$N_b=2400$ 
and light grains, with $M_s=0.03$ g and $\phi_s=0.41$, %$N_s=2700$
all with the same diameter $D=1$ cm, prepared in a horizontally fully segregated configuration (Fig. 1(a)).
The mixture is vibrated on an horizontal tray along the x-direction with amplitude $A$ and frequency $\nu$. 
\begin{figure*}[t!!]
\includegraphics*[scale=1]{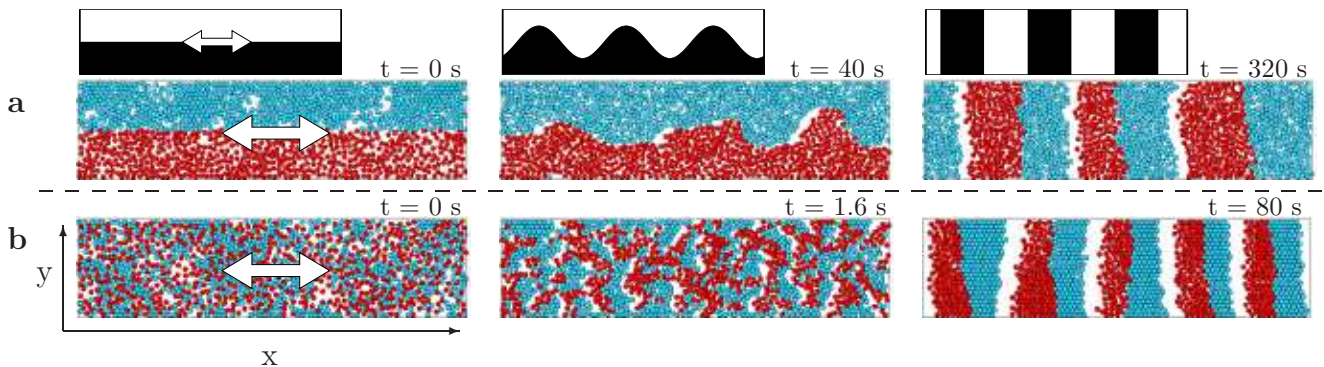}
\caption{\label{fig-modello}
(color on-line) Evolution of a binary mixture of heavy (red-dark gray) and light (blue-ligth gray) grains on an horizontal tray oscillating along the $x$-direction with amplitude $A=1.2$~cm and frequency $\nu=12$~Hz (only a quarter of the system length is shown in each figure). 
{\bf Panels a} The system is initially prepared in a horizontally segregated configuration with a flat interface between the two species. Under shaking it develops a surface instability with a sine like modulation of growing amplitude leading at long times to a state where segregated stripes perpendicular to the driving direction appear. 
{\bf Upper small panels} show the corresponding evolution of the surface instability model Eq.s~(\ref{model1})
in presence of an oscillating shear $\Delta v(t)=2\pi A\nu\sin(2\pi\nu t)$.
{\bf Panels b} The same system is now initially prepared in a mixed state and then shaken.
The formation of local fluctuations of the density of the two species (shown at $t=1.6$~s) leads to microsurfaces generating the same instability mechanism as in panels a. 
%The instability grows in this case too, giving rise to a similar pattern of segregated stripes as seen before (shown at $t=80$ s), even though the final number of stripes can be different.
}
\end{figure*}
The characteristic time scale, $\tau_b=M_b/\mu_b$ and $\tau_s=M_s/\mu_s$, is different for the two species and they are forced to oscillate with different amplitudes and phases, subject to an effective surface shear periodically varying in time. 
%Under this mechanism, the initially flat interface between the two components is observed to evolve (see Fig. 1 a) via the formation of a surface modulation which has a growing amplitude.
One may expect configurations in which the two different components form two stripes parallel to the driving direction (see Fig. 1(a)) to be stable, the two stripes oscillating independently with different amplitude and phases. Instead the initially flat interface between the two components is observed to evolve via the formation of a surface modulation which has a growing amplitude. Finally, this leads to a state where segregated stripes appear, perpendicular to the driving direction (at variance with known results in colloidal fluids \cite{colloidi}).
This instability appears of the same kind of that observed in two fluid systems and in liquid sand systems subject to horizontal oscillations under gravity, which is responsible for the ripples observed on the shoreline \cite{Scherer1999}. In these cases, however, because of gravity there is an energy cost associated to the growth of the interface, which therefore gets stabilized.

%Interestingly, when the system starts from a disordered initial state, as shown in Fig. 1 b, the instability develops too, via the formation of local fluctuations of the density of the two species; the originally disordered system is ordered, i.e., segregated, by the dynamic instability. 

%% modifica aggiunta per giustificare la instability induced segregation
When the system starts from a disordered initial state, as shown in Fig. 1(b), the instability develops too. The formation of local fluctuations of the density of the two species create microsurfaces which are forced to evolve according to the instability shown in Fig. 1(a). For instance Fig. 1(b), at t$=1.6$~s, shows the existence of clusters of particles of the same species elongated perpendicularly to the driving direction.
As time goes on these clusters merge giving rise to the striped pattern of Fig. 1(b) (t$=80$~s). It is apparent that the originally disordered system is ordered, i.e., segregated, by the dynamic instability. 
Fig. 1(a) (t$=320$~s) and Fig. 1(b) (t$=80$~s) show that the wavelength of the final steady state depends on the initial conditions.

The segregation process shown in Fig. 1(a) reproduces that observed experimentally in \cite{Mullin2000, Reis2004}. In the present set-up grains radii are equal, no ``depletion'' forces can be at work, and no ``thermodynamic'' coarsening phenomena can be considered responsible for this segregation process, as initially conjectured \cite{Reis2004} by making an analogy with colloidal systems.
Such a remark appears to be consistent with a scenario proposed by use of modified Navier-Stokes equations \cite{Yeomans}. Interestingly, segregation from a disordered initial state to a final striped one has also been observed in a two fluid system oscillated in absence of gravity \cite{Wunenburger2000}.

The above dynamical instabilities should manifest in the mixture also under different shearing protocols. To check this, we study now a different configuration where gravity induces granular flow and thus a differential shear on the mixture components appears. 
The tray no longer oscillates, but it is inclined of an angle $\theta$ with the respect to the horizontal by rotation around the y-axis.The two species of grains attain different limit velocities (of the order of
$v_b\simeq g\tau_b\sin\theta$ and $v_s\simeq g\tau_s\sin\theta$, with $g$ gravitational acceleration) and thus
experience a differential shear. As shown in Fig. 2, under these conditions the initially flat interface between the mixture components changes too, via Kelvin-Helmholtz-like instabilities. An unstable pattern forms
and curl-like structures appear. This phenomenon, induced by surface dynamical instabilities, is very similar to those recently discovered in experiments on granular flows \cite{Goldfarb2002}.

\begin{figure*}[t!!]
\psfig{figure=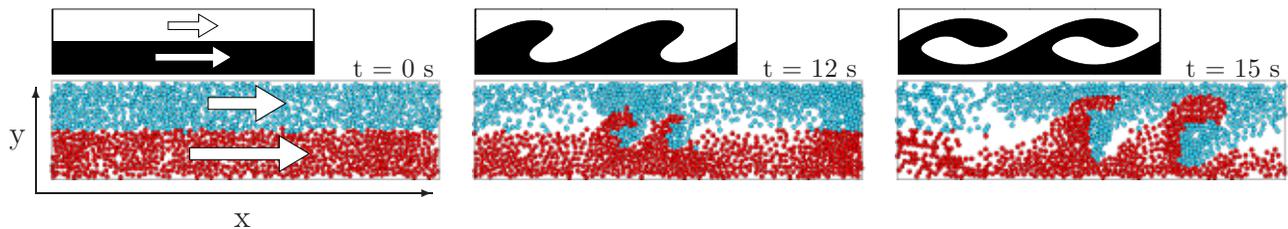,scale=1,angle=0}
\caption{
(color on-line) Evolving interface in the same granular mixture of Fig.~1a 
flowing down a tray rotated of an
angle $\theta=12^o$ with respect to the horizontal around the y-axis.
The red (dark gray) grains are faster that the blue (light gray) ones.
%The system is initially prepared in a fully segregated
%state ($t=0$ s). As time goes on, the straight interface develops
%a pattern, as a surface instability induced by differential shear
%appears between the two components.
{\bf Upper small panels}
The upper panels show the evolution, at corresponding times, of a
Kelvin-Helmholtz-like surface dynamical instability modeled by
Eq.s~(\ref{model1}) in presence of a constant shear $\Delta v(t)=\Delta v_0$,
reproducing a pattern qualitatively similar to the one found in the
main panels.
%Eq.s~(\ref{model1}) was originally proposed to describe surface shear
%instabilities in granular flows \cite{Goldfarb2002}, which appear to be of
%the same kind of the present ones.
}
\label{fig-interface}
\end{figure*}

The results of Fig. 1 and Fig. 2 point out the essential role of shear
induced surface instabilities in pattern formation and the connections
with segregation transitions. Even though this is not an usual fluid, 
the complex nature of this phenomenon can be understood via schematic
hydrodynamic-like considerations.
On the two sides of the perturbed interface flux lines are
narrowed and widened as the pressure is respectively lowered and increased
(Bernoulli law). The pressure gradient reinforces the perturbation
of the interface giving rise to a positive feedback.
Within such a perspective, the essential features of pattern formation observed
in the two experiments of Fig.s 1 and 2 are captured by a
model originally proposed for the Kelvin-Helmholtz instability of
granular flows \cite{Goldfarb2002}.
The model describes the evolution of the velocity components,
$v_X$ and $v_Y$, of the system interface initially located at $Y(X) = 0$:
\begin{equation}
v_X = \Delta v(t) \tanh(Y) ~~~~~
v_Y = c \sin(k_X X)
\label{model1}
\end{equation}
where $\Delta v(t)$ is the relative bulk velocity between grains far on the
two side of the interface at time $t$, as $k_X$ and $c$ are constant
parameters (we consider the same values used in \cite{Goldfarb2002}, i.e.,
$k_X = 5$, $c = 0.1$, but our results are robust to changes).
The presence of a velocity gradient, $\Delta v(t)$, independently
from its microscopic origin (different grains friction, different driving
of the components, etc...), enhances shear and results in patterns formation.
In the case of gravity induced deformations,
$\Delta v(t)=\Delta v_0\equiv v_b-v_s$ can be considered to be time
independent since in the stationary regime the two limit velocities are given.
In such a case, the model essentially coincides with the one
introduced in \cite{Goldfarb2002} and well describes the qualitative
features of the MD simulations of Fig. 2 (see upper panels).
In the case of the vibrated horizontal tray, the relative bulk
velocity is oscillating in time and we fix
$\Delta v(t) = 2\pi A\nu\sin(2\pi\nu t)$.
This gives rise (see upper panels in Fig.1(a)) to a growing interface
with features very close to those observed in the simulations.

\textit{Pattern formation --}
We show now under which conditions pattern formations occur and
explain how the properties of the final state of the mixture,
such as the emerging characteristic length scale of the stripes, $\lambda$,
depend on both the dynamics control parameters and the relative
concentration of the two components.
This gives rise to a complex mixing/segregation diagram which we discuss
in details here in the case of the horizontally vibrated experiment.
In order to make direct comparisons with experimental results
on segregation observed by Mullin and coworkers
\cite{Mullin2000,Reis2002,Reis2004,Ehrhardt2004},
we consider now a binary mixture of
large mono-disperse disks, of diameter $D_b=1$cm, covering an
area fraction $\phi_b$, and small polydisperse disks~\cite{nonpoly}, with average
diameter $D_s=0.7$cm (with $17$\% polydispersity) and area fraction $\phi_s$. 
Masses and frictional parameters are as given before. 

The system starts from a mixed initial configuration as in Fig. 2(b).
The nature of the state reached by the system under shaking at late stage
is crucially dependent on the area fraction of the two species:
segregation in stripes is only found for high enough concentrations.
This behaviour, in the ($\phi_b,\phi_s$) plane, is summarized in the diagram
of Fig. 3(a) showing the system ``fluid'' and ``crystal''
regions along with their segregation properties, for $\nu=12$ Hz and $A=1.2$ cm.
Large grains are considered to be in a ``fluid'' configuration when
their radial density distribution function, $g(r)$,
shows a first peak at $r = D_b$ and a second one at $r=2D_b$, and to be
in a ``crystal'' configuration when a new peak at $r =\sqrt{3}D_b$
appears \cite{Reis2004}. The system is in a ``glassy'' state \cite{Capri} when on the longest
of our observation time scales, the system is still far from stationarity.

\begin{figure}[t!!]
\includegraphics*[scale=0.9]{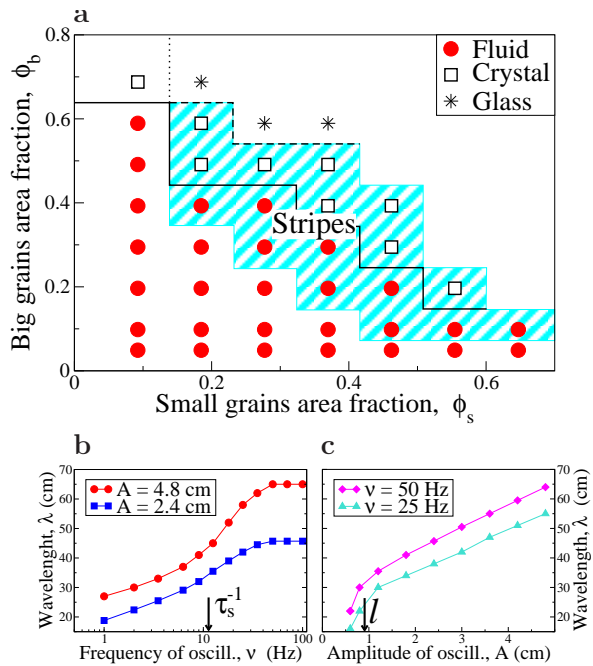}
\caption{\label{fig-diagram-schematic}
{\bf Panel a} Ordering properties of the late stage
configurations of the  mixture as a function of the area fractions of the two components.
The {\em shaded area} covers the region where segregation via stripes formation occurs.
{\bf circles}: large grains are in a fluid state.
{\bf squares}: large grains form a crystal. 
{\bf stars}: the system appears blocked in a ``glassy''
disordered configuration (see text).
%The system species at small concentrations are mixed and in a fluid
%state ({\bf circles}). When the concentration increases, segregation
%via stripes formation appears ({\em shaded region} named
%{\bf ``Stripes''}). At higher concentrations, large grains have a transition
%to a crystal state ({\bf squares}). Smaller grains are always in a
%fluid configuration because of their polydispersity.
%Finally, at very high area fraction the system appears blocked in a ``glassy''
%(see text) disordered configuration ({\bf stars}).
When stripes form their characteristic length scale $\lambda$
is a function of the frequency, $\nu$, and of the amplitude, $A$,
of the driving oscillations. This is shown, in the case
$\phi_b=0.30$ and $\phi_s=0.28$, in {\bf panels b} and {\bf c}.
}
\end{figure}

Fig. 3(a) shows that grains at small concentrations are mixed and in a
fluid state. Segregation via stripes formation appears at higher
concentrations. At even higher concentrations, large grains form stripes
with a crystalline order, as smaller grains are always fluid for their
polydispersity. Finally, at very high area fractions, the system is
blocked in its starting disordered configuration (``glassy'' region).
For instance, by increasing $\phi_s$ at a fixed value of $\phi_b$
(say $\phi_b \simeq 0.174$), we observe first a transition from a mixed
fluid state to a segregated striped fluid and then a transition where the
the monodisperse phase crystallize. The experiments of \cite{Reis2004},
where $\phi_b \simeq 0.174$, show the very same transitions found here
at locations differing by a $10\%$.
%By increasing the area fraction of one component we observe first segregation of the system 
%via the formation of stripes, and then crystallization of the monodisperse phase.
%The experiments \cite{Mullin2000,Reis2002,Reis2004} are done at a fixed value of the area fraction of the monodisperse phase
%($\phi_b \simeq 0.174$). For this value of $\phi_b$ our results of Fig.~3a differ of about $10\%$.

%The experiments in \cite{Mullin2000,Reis2002,Reis2004}, made in packings with a given area fraction of large grains
%(i.e, given $\phi_b$), clearly indicates that by increasing the fraction of small grains ($\phi_s$) 
%the system has two ``transitions'': a first one where stripes start forming as the system is still all ``fluid''; a second
%one where stripes of large grains appear to crystallize. This is precisely the behaviour we find, although
%the location of these transitions does not perfectly coincide with that reported in the experiment of Ref. \cite{Reis2004} (deviation about 10\%).

In the case $\phi_b=0.30$ and $\phi_s=0.28$, where stripes
form, we describe their dependence on the dynamics control parameters
in Fig.s 3(b) and 3(c), showing that the length scale, $\lambda$, increases
as a function of the shaking frequency, $\nu$, and of the amplitude $A$.
These results are to be compared, for instance, with those found in
liquid-sand mixtures under oscillating flow, where ripples form with a wavelength depending
on the amplitude of oscillation, but not on its frequency \cite{Stegner1999}.
The dependence on $\nu$ can be schematically understood by comparison
with the characteristic time scales
$\tau_b=M_b/\mu_b$ and $\tau_s=M_s/\mu_s$ of the two species
(here $\tau_b^{-1}=0.28$ Hz and $\tau_s^{-1}=11.3$ Hz):
in the limit $\nu \gg \tau_b^{-1},\tau_s^{-1}$
grains are not able to follow the tray motion and no sensitivity to
$\nu$ is expected, as well as when $\nu \ll \tau_s^{-1}$
since the grains move with the tray.
Analogously, the dependence on $A$ is expected to be substantial when
$A$ is at least of the order of the mean grains separation length,
$l=(4\phi_b/\pi D_b^2+4\phi_s/\pi D_s^2)^{-1/2}$,
since under this condition grains strongly interact.

In conclusion our simulations of a binary mixture under horizontal vibrations 
have revealed the existence of a new dynamical instability and shed light 
on the process of size segregation under oscillatory shear and its 
connections to pattern formation in granular flows.
Within such a unifying framework, we derived the ``phase diagram'' 
of the mixing/segregation states of the mixture and its corresponding 
transitions. Finally, such hydrodynamic like processes appear to be related 
to those known in thermal fluids and fluid-grains systems, 
ranging from Kelvin-Helmholtz instabilities to ripples formation 
in liquid sand mixtures, even though interesting difference are found.

Work supported by EU Network MRTN-CT-2003-504712, MIUR-PRIN 2004, MIUR-FIRB 2001, CrdC-AMRA, INFM-PCI.


\begin{thebibliography}{40}
\bibitem{nonpoly}
If both species are monodisperse we do observe a phase diagram which is similar to that of Fig 3. However when segregated
in stripes both species can be found in a liquid or in a crystal-like configuration.
\end{thebibliography}
\end{document}